\documentclass[12pt]{article}

\usepackage{amsmath}
\usepackage{cite}
\usepackage{epsfig}

\begin{document}

\rightline{CERN-PH-TH/2009-095}

\begin{center}
{\Large\bf Rapidity and energy dependence of average}\\
\vskip 0.3cm
{\Large\bf  transverse momentum and particle density}\\
\vskip 0.3cm
{\Large\bf in saturation models}
\vspace{0.8cm}

P.~Brogueira$^{1}$, J.~Dias~de~Deus $^{2}$, and  J.~G.~Milhano$^{2,3}$

\vspace{0.2cm}
 
{\it $^1$ Departamento de F\'isica, Instituto Superior T\'ecnico (IST) ,\\
Av. Rovisco Pais, P-1049-001 Lisboa, Portugal}
\vskip 0.1cm
{\it $^2$ CENTRA, Departamento de F\'isica, Instituto Superior T\'ecnico (IST),\\
Av. Rovisco Pais, P-1049-001 Lisboa, Portugal
}
\vskip 0.1cm
{\it $^3$ Theory Group, Physics Department, CERN,\\
CH-1211 Gen\`eve 23, Switzerland
}

\end{center}
\vskip 0.3cm
{\small
Saturation models -- colour glass condensate and string percolation -- impose a strict relation between the average transverse momentum, $\langle P_T\rangle$, and the rapidity particle densities, $dn/dy$. By combining this relation with an appropriate evolution equation for $dn/dy$, and imposing energy-momentum conservation, we obtain a fair description of data, for generic $AB$ collisions (hadron-hadron, hadron-nucleus and nucleus-nucleus) at all rapidities and (high) energies. Predictions are given for the LHC. 
}


\section{Introduction}
\label{sec:intro}

Over the last few years the relationship between the phenomenological successful String Percolation Model (SPM) \cite{Armesto:1996kt} and the theoretically sound Colour Glass Condensate (CGC) approach \cite{McLerran:1993ni,McLerran:1993ka,McLerran:1994vd,Jalilian-Marian:1996xn,Jalilian-Marian:1997dw,Kovner:1999bj,Kovner:2000pt,Weigert:2000gi,Iancu:2000hn,Ferreiro:2001qy,Weigert:2005us} to high energy hadronic processes has been clarified \cite{Kharzeev:2006zm,DiasdeDeus:2006xk,Armesto:2006bv,DiasdeDeus:2007wb,Dias de Deus:2007tf,Brogueira:2007ub}.
Such a relationship provides not only much needed theoretical justification for the assumptions underlying the SPM, but also a simplified framework in which to perform phenomenological studies accounting for the essential features of the CGC.

In \cite{DiasdeDeus:2007wb} it was argued that complementing the SPM by a simple evolution equation --- which describes the generation of lower rapidity strings from higher rapidity ones --- yields an effective description of the Glasma \cite{Lappi:2006fp,Romatschke:2005pm,Romatschke:2006nk,Lappi:2006nx,Lappi:2006hq,Gelis:2006ks}, i.e. of the highly coherent state formed immediately after the collision of two CGC sheets. Thus, the SPM incarnation of \cite{DiasdeDeus:2007wb} can be used to address the early post-collision dynamics in a heavy ion collision. A series of predictions  \cite{Abreu:2007kv,DiasdeDeus:2007tf,Brogueira:2007ub}, in which RHIC data was used to fix the model parameters, for observables at the forthcoming LHC Pb-Pb and p-Pb programmes have been obtained in the framework of this model.
Importantly, the results obtained for Forward-Backward correlations \cite{Brogueira:2007ub} were found to be in agreement with those obtained in the CGC/Glasma \cite{Armesto:2006bv}.

In the present work we address the effects due to the inclusion of average transverse momentum in our model. From the observation that saturation models --- CGC and SPM --- impose a strict relation between average transverse momentum and particle density distributions in rapidity, we obtain, imposing energy momentum conservation \cite{Dias de Deus:2007tf}, both particle density distributions and average transverse momentum distributions in rapidity.

The paper is organized as follows. In Sec. \ref{sec:model} we describe our model. Our results, including both a comparison with available RHIC data and predictions for the LHC, are presented in Sec. \ref{sec:results}. Sec. \ref{sec:conclusions} contains our conclusions.

\section{Model setup}
\label{sec:model}

The SPM \cite{Armesto:1996kt} deals, as its fundamental degrees of freedom, with the longitudinally extended strings formed in the collision.
The relevant parameter to address percolation phenomena is the density $\xi$\,\footnote{We will denote the density by $\xi$, instead of the traditional $\eta$ in order to avoid confusion with the pseudo-rapidity variable.} which, in the impact parameter plane, is defined as
\begin{equation}
\label{eq:transdens}
	\xi \equiv \bigg(\frac{r}{R}\bigg)^2 N_s\, ,
\end{equation}
where $N_s$ is the number of strings, or discs in the impact parameter plane, $r$ is the radius of each disc, and $R$ is the effective radius of the interacting overlapping area. 
For an $AB$ collision, with $A\leq B$, the area of overlap is determined by $A$ or, more precisely, by the number of participating nucleons from $A$, i.e. by $N_A$ ($1\leq N_A \leq A$).
In the following we shall make the natural  and simplifying assumption that
\begin{equation}
\label{eq:rad}
	R\simeq R_P N_A^{1/3}\, ,
\end{equation}
where $R_P$ is the nucleon radius.

The average number of collisions $N_{\rm col}$ can be estimated given the numbers $N_A$ and $N_B$ of participating nucleons from each nucleus. As the overlap area is given by $N_A^{2/3}$, the $N_B$ participating nucleons from $B$ are the ones crossing that area, such that each of them interacts with $N_A^{1/3}$ nucleons from $A$. We thus have
\begin{equation}
\label{eq:ncol}
N_{\rm col} \simeq N_A^{1/3} N_B\, .
\end{equation}
When $A=B$, (\ref{eq:ncol}) reduces to the well known Glauber result and for $N_A = 1$,  $N_{\rm col} = N_B$ as one would also expect.

Two important observations are essential for our argument: 

(i) There is a definite relation between the averaged transverse momentum $\langle p_T \rangle$ and the rapidity particle distribution $dn/dy$ \cite{McLerran:2001cv,SchaffnerBielich:2001qj}. The existence of such a relation is related to the presence of a single scale in the problem, be it the saturation scale in the case of the CGC,  or, in the case of Schwinger strings, that the field and the charge are related to each other by  Gauss' theorem.

(ii) The appearance of low rapidity objects (gluons or strings) arising from fast coloured objects (glasma fields or valence strings) can be described by an evolution equation \cite{Lappi:2006fp,Romatschke:2005pm,Kharzeev:2006zm,DiasdeDeus:2007wb}.

\subsection{The relation between  $\langle p_T \rangle$ and $dn/dy$}
\label{sec:ptdndy}

The occurrence of percolating behaviour depends on whether strings do or do not overlap.  
If there is no overlap of strings, percolation does not occur and one expects the average transverse momentum $\langle p_T^2 \rangle$ to be given simply by the average transverse momentum ${\bar p}_{1}^{2}$ associated with production from a single string,
\begin{equation}
	\langle p_T^2 \rangle = {\bar p}_{1}^{2}\quad \mbox{and} \quad \frac{dn}{dy} = N_{s} {\bar n}_{1}\, ,
\end{equation}
where ${\bar n}_{1}$ is the particle density associated with a single string and $N_{s}$ is the number of strings.

On the other hand, when strings overlap, percolation will occur and colour vector summation will result both in a suppression of the overall colour charge and in an enhancement of the effective string tension, i.e.
\begin{equation}
\label{eq:dndy1}
	\frac{dn}{dy} = F(\xi ) N_s {\bar n_1} \, ,
\end{equation}
which can be rewritten, using (\ref{eq:transdens}) and (\ref{eq:rad}), as
\begin{equation}
\label{eq:dndy2}
\frac{1}{N_A^{2/3}} \frac{dn}{dy}  = F(\xi) \xi \bigg(\frac{R_{P}}{r}\bigg)^2 \bar n_1 \, .
\end{equation}
For the average transverse momentum we get
\begin{equation}
\label{eq:pt2}
	\langle p_T^2\rangle = \frac{{\bar p_1}^2}{F(\xi)} \, .
\end{equation}
In both (\ref{eq:dndy2}) and (\ref{eq:pt2}), $F(\xi)$ is the colour suppression factor given by \cite{Braun:2001us}
\begin{equation}
\label{eq:fxi}
	F(\xi) \equiv \sqrt{\frac{1-e^{-\xi}}{\xi}}\, .
\end{equation}

The only varying quantity in the right hand side of both (\ref{eq:dndy2}) and (\ref{eq:pt2}) is the transverse density $\xi$. Thus, a direct relation between  $1/N_A^{2/3}\cdot {dn/dy}$ and $\langle p_T^2\rangle$ necessarily exists \cite{McLerran:2001cv,SchaffnerBielich:2001qj,DiasdeDeus:2003fg,DiasdeDeus:2005yt}. In fact, as $F(\xi)$ is a decreasing function of $\xi$ and $F(\xi) \xi$ is an increasing function of $\xi$,  there is a monotomic functional dependence between $dn/dy$ and $\langle p_T^2\rangle$. This result is essential in our work.

Eq. (\ref{eq:pt2}) for the average transverse momentum $\langle p_T^2\rangle$  does not account for any kinematical effects. As $\xi \to 0$, $F (\xi) \to 1$ and  (\ref{eq:pt2}) yields $\langle p_T^2 \rangle \to {\bar p_1}^2$ (the average transverse momentum of a single string), whereas one would expect that in this limit and for a particle of mass $m$ and with rapidity  $y_m = \ln \sqrt{s/m}$, where $\sqrt{s}$ is the centre of mass energy, the averaged transverse momentum $\langle p_T^2\rangle$, at the end of phase space, should be zero. Thus, eq.  (\ref{eq:pt2}) needs to be kinematically corrected to the form
\begin{equation}
\label{eq:pt2corr}
	\langle p_T^2 \rangle = \frac{{\bar p_1}^2}{F(\xi)} \frac{1}{\big[1-\big({s_0/s}\big)^a\big]+\big({s_0/s}\big)^a \cosh^2 \eta}\, ,
\end{equation}
where $\eta$ is the pseudo-rapidity. The parameters  $s_0$ and $a$ in (\ref{eq:pt2corr}) are fixed by the standard relation between rapidity and pseudo-rapidity, taken in the limit $y\to y_m$ and $\eta \to \infty$, to
\begin{equation}
\label{eq:parfix}
	s_0 = 4{\bar p_1}^2 \, ,\quad a=1\, ,
\end{equation}
such that 
\begin{equation}
\label{eq:pt2corrfin}
	\langle p_T^2 \rangle = \frac{{\bar p_1}^2}{F(\xi)} \frac{1}{\big[1-\big({4{\bar p_1}^2/s}\big)\big]+\big({4{\bar p_1}^2/s}\big) \cosh^2 \eta}\, .
\end{equation}
The ratio $s_0/s$ is, for high enough energies, very small. Thus, the correction (\ref{eq:pt2corr}) to (\ref{eq:pt2}) is only important for very large values of $\eta$.

In the high density region $\xi \gg 1$ corresponding, as we shall see, to mid-rapidity, the number of produced strings is proportional to the number of collisions, i.e. $N_s \sim N_{\rm col}$ \cite{DiasdeDeus:2000cg,DiasdeDeus:2000gf}. Then, from (\ref{eq:transdens}), (\ref{eq:rad}) and  (\ref{eq:ncol}), it follows that
\begin{equation}
\label{eq:chidense}
	\xi \sim N_A^{-1/3} N_B\, ,
\end{equation}
and that, from (\ref{eq:dndy1}) (or (\ref{eq:dndy2})), 
\begin{equation}
\label{eq:dndydense}
	\frac{dn}{dy} \sim N_A^{2/3} \xi^{1/2} \sim N_A^{1/2} N_B^{1/2}\, .
\end{equation}

In the low density region $\xi \ll 1$ corresponding to the forward (backward) rapidity region, one has $N_s \sim N_A$ ($N_s \sim N_B$)  \cite{DiasdeDeus:2000cg,DiasdeDeus:2000gf}. Thus, we have
\begin{alignat}{2}
	\xi &\sim N_A^{1/3}  \quad &\mbox{(Forward)}\, ,\label{eq:xiforward}\\
	\xi &\sim N_A^{-2/3} N_B  \qquad &\mbox{(Backward)}\, ,\label{eq:xibackward}
\end{alignat}
and
\begin{alignat}{2}
	\frac{dn}{dy} &\sim N_A  \qquad &\mbox{(Forward)}\, ,\label{eq:dndyforward}\\
	\frac{dn}{dy} &\sim N_B  \qquad &\mbox{(Backward)}\, .\label{eq:dndybackward}
\end{alignat}

For symmetric collisions $N_A = N_B$, we obtain, as expected,  $dn/dy \sim N_A$, asymptotically, for all rapidities.
However, in the asymmetric case $N_A \neq N_B$, (\ref{eq:dndyforward}) and 
(\ref{eq:dndybackward}) do not accurately account for the relevant physics. In fact, what is implicit in (\ref{eq:dndyforward}) and (\ref{eq:dndybackward}) is that nucleons interact in pairs and that each nucleon only interacts once, so that particle production is proportional to that from $pp$ with a proportionality factor $N_{A}$ ($N_{B}$).
When $N_A < N_B$ at least some of the nucleons from $A$ reinteract with nucleons from $B$ being decelerated relatively to the result of (\ref{eq:dndyforward}). Correspondingly, in the  backward direction the nucleons from $B$ are less decelerated 
than the result from  (\ref{eq:dndybackward}). In conclusion, $dn/dy$ is overestimated in  (\ref{eq:dndyforward}) and underestimated in (\ref{eq:dndybackward}). We account for this by introducing corrections factors $(N_A/N_B)^\gamma$ and $(N_B/N_A)^\gamma$ in 
(\ref{eq:dndyforward}) and (\ref{eq:dndybackward}) respectively. Thus, we have
\begin{alignat}{2}
	\frac{dn}{dy} &\sim N_A  \bigg(\frac{N_A}{N_B}\bigg)^\gamma\qquad &\mbox{(Forward)}\, ,\label{eq:dndyforwardcorr}\\
	\frac{dn}{dy} &\sim N_B   \bigg(\frac{N_B}{N_A}\bigg)^\gamma\qquad &\mbox{(Backward)}\, .\label{eq:dndybackwardcorr}
\end{alignat}

Clearly, (\ref{eq:dndyforward}) and (\ref{eq:dndybackward}) are recovered in the limit $N_{A}=N_{B}$. The phenomenological parameter $\gamma$ is to be adjusted. The effect encoded in (\ref{eq:dndyforwardcorr}) and (\ref{eq:dndybackwardcorr}) has been observed \cite{Barton:1982dg,pajaresbook} in forward production of $\pi^\pm$ in $pp$ collisions with  $\gamma \simeq 0.11$.

\subsection{The evolution equation for  $dn/dy$}
\label{sec:evol}

We will use the framework developed in \cite{DiasdeDeus:2007wb} to describe the generation of low rapidity strings from a high forward rapidity (valence) string. In a nutshell, this generation is given by the logistic equation for population dynamics
\begin{equation}
\label{eq:evol}
	\frac{\partial \rho}{\partial (-\Delta )} = \frac{1}{\delta} \rho \bigg(1-\frac{\rho}{\rho_{Y}} \bigg)\, ,
\end{equation}
where $\rho \equiv \rho (\Delta , Y) \equiv dn/dy$ is the particle density, $Y\equiv \ln (\sqrt s /m_{b})$ is the beam rapidity and
\begin{equation}
\label{eq:rapdif}
	\Delta \equiv \eta - Y\, ,
\end{equation}
with $\eta$ the pseudo-rapidity.
The  variable $-\Delta$ plays the role of evolution time. The parameter $\delta$ controls the low density evolution of $\rho$ and must therefore depend on intrinsic parameters of the theory. $\rho_Y$ is the asymptotic saturation density.

The solution of (\ref{eq:evol}) must be such that 
\begin{equation}
	\frac{\partial \rho}{ \partial (-\Delta)} = 0
\end{equation}	
both at  $\rho =0$ and $\rho =\rho_Y$, and 
\begin{equation}
	\frac{\partial^2 \rho}{ \partial (-\Delta)^2}\bigg|_{\Delta_0}  = 0\, ,
\end{equation}	
where the scale $\Delta_0$ defines the separation between the region $\Delta > \Delta_0$, of low density and positive curvuture, and the region $\Delta < \Delta_0$ of high density and negative curvuture, and is such that $\rho_0 \equiv \rho (\Delta_0 , Y) =\rho_Y/2$.
The integration of (\ref{eq:evol}) yields
\begin{equation}
\label{eq:sol}
	\rho (\Delta ,Y) = \frac{\rho_Y}{e^{\frac{\Delta -\Delta_0}{\delta}} + 1} \, ,	
\end{equation}
which is nothing but a generalization of the Fermi distribution, known to approximately fit RHIC data for pseudo-rapidity distributions in central nucleus-nucleus collisions \cite{Adams:2005cy,Brogueira:2006nz}. In \cite{Brogueira:2006nz} it was argued that 
\begin{equation}
\label{eq:rhoy}
	\rho_Y =e^{\lambda Y}\, ,	
\end{equation}
and
\begin{equation}
\label{eq:delta0}
	\Delta_0 =- \alpha Y \, ,	
\end{equation}
where $\alpha$ and $\lambda$ are positive constants and $\alpha, \lambda \leq 1$. The distribution (\ref{eq:sol}) is then characterized by three constants: $\alpha ,\delta$ and $\lambda$.

In a collision one has not only forward emission, but also backward emission. For symmetrical situations, (\ref{eq:sol}) describes both forward emission ($\eta \geq 0$) and backward emission  ($\eta \leq 0$). An analogous line of reasoning underlies related approaches in the context of the CGC \cite{Kharzeev:2004if}.

In the most general situation, with $N_A$ participant nucleons moving in the forward direction and $N_B$ participants moving in the backward direction, the fulfilment of (\ref{eq:dndydense}), (\ref{eq:dndyforwardcorr}) and (\ref{eq:dndybackwardcorr}), leads to 
\begin{equation}
\label{eq:forwasym}
	\frac{dn}{dy} = \frac{N_A^{1/2} N_B^{1/2} e^{\lambda Y}}{(N_B/N_A)^{1/2+\gamma}\, e^{\frac{\eta-(1-\alpha) Y}{\delta}}+1} \qquad \mbox{(F)}\, ,
\end{equation}
and 
\begin{equation}
\label{eq:backasym}
	\frac{dn}{dy} = \frac{N_A^{1/2} N_B^{1/2} e^{\lambda Y}}{(N_A/N_B)^{1/2+\gamma}\, e^{\frac{-\eta-(1-\alpha) Y}{\delta}}+1} \qquad \mbox{(B)}\, .
\end{equation}
Note that in general the forward (F) and backward (B) distributions meet at $\eta_c=\delta(\frac{1}{2} +\gamma) \ln \frac{N_A}{N_B} \leq 0$.

If $N_A = N_B$, (\ref{eq:forwasym}) and (\ref{eq:backasym}) become mirror distributions with, for the forward region,
\begin{equation}
\label{eq:forsym}
	\frac{dn}{dy} = \frac{N _A \, e^{\lambda Y}}{e^{\frac{\eta -(1-\alpha)Y}{\delta}}+1}\, .
\end{equation}

In order to allow for direct comparison with experimental data, the multiplicity distributions --- (\ref{eq:forsym}) for symmetric collisions and (\ref{eq:forwasym}) and (\ref{eq:backasym}) for asymmetric ones --- must be rewritten as distributions in pseudo-rapidity $\eta$. Performing the standard transformation the multiplicity distribution in the symmetric case is given by 
\begin{equation}
\label{eq:pseudoforsym}
	\frac{1}{N_A}\frac{dn}{d\eta} = J	\,
	\frac{e^{\lambda Y}}{e^{\frac{\eta -(1-\alpha)Y}{\delta}}+1}\, ,
\end{equation}
For an asymmetric collision we obtain
\begin{equation}
\label{eq:pseudoforwasym}
	\frac{1}{(N_A  N_B)^{1/2}}\, \frac{dn}{d\eta} 
	= J \, 
	\frac{e^{\lambda Y}}{(N_B/N_A)^{1/2+\gamma}\, 
	e^{\frac{\eta-(1-\alpha) Y}{\delta}}+1} 
	\qquad \mbox{(F)}\, ,
\end{equation}	
and
\begin{equation}
\label{eq:pseudobackasym}
	\frac{1}{(N_A  N_B)^{1/2}}\,
	\frac{dn}{d\eta} = J \,
	\frac{e^{\lambda Y}}{(N_A/N_B)^{1/2+\gamma}\, 
	e^{\frac{-\eta-(1-\alpha) Y}{\delta}}+1} 
	\qquad \mbox{(B)}\, ,
\end{equation}
In both cases the transformation Jacobian is given by
\begin{equation}
\label{eq:jac}
	J= \frac{dy}{d\eta} = \frac{\cosh \eta}{\sqrt{k+ \sinh^2\eta}}\, ,\quad k=\frac{m^2+p_T^2}{p_T^2}\, .
\end{equation}

\subsection{Energy-momentum conservation}
\label{sec:emcons}

As we are dealing with $dn/dy$ distributions for all rapidities, or pseudo-rapidities, and as we have some control on the $p_T$ distribution, via $\langle p_T^2\rangle$, it is natural to impose, in some simplified scheme \cite{Dias de Deus:2007tf}, energy-momentum conservation.

We write for the  overall energy $E$ conservation
\begin{equation}
\label{eq:econs}
	E= \int_{-y_m}^{+y_m} \langle m_T\rangle \cosh y \frac{dn}{dy} dy = (N_A + N_B) \frac{\sqrt{s}}{2}\, ,	
\end{equation}
and for the overall longitudinal momentum 
\begin{equation}
\label{eq:pcons}
	{\vec P} = \int_{-y_m}^{+y_m} \langle m_T\rangle \sinh y \frac{dn}{dy} dy= (N_A -N_B) \frac{\sqrt{s}}{2}\, ,
\end{equation}
where the transverse effective mass $\langle m_T\rangle$ is given by
\begin{equation}
\label{eq:mt}
	\langle m_T\rangle \equiv \sqrt{\langle p_T^2\rangle +m^2}\, ,
\end{equation}	
with  $m$ an effective mass (introduced to avoid the sum over all species).

In our comparisons with data we impose (\ref{eq:econs}) and (\ref{eq:pcons}). Further, in order to go from the experimentally measured $dn/d\eta$ (charged) to the $dn/d\eta$ (total), we use a multiplying factor, consistent with dominance of $\pi^+,\pi^-, \pi^0$ production,  which is of the order of $3/2$. 
It is, however, possible to construct, from  (\ref{eq:econs}) and (\ref{eq:pcons}), a ratio which is independent  of multiplying factors
\begin{equation}
\label{eq:peratio}
	\frac{| \vec P |}{E} = \frac{N_B - N_A}{N_A +N_B} \, .
\end{equation}

\section{Results: comparison with experimental data}
\label{sec:results}

The model parameters $\alpha$, $\delta$ and $\lambda$ in the multiplicity distributions for both the symmetric case (\ref{eq:pseudoforsym}) and for the asymmetric one, (\ref{eq:pseudoforwasym}) and (\ref{eq:pseudobackasym}), along with the parameter $\gamma$ (only relevant for asymmetric collisions) were fixed as to provide the best model joint description of RHIC data ($\sqrt{s}=200$ GeV)  for Au-Au central collisions and d-Au collisions in several centrality classes \cite{Back:2004je}, and to further satisfy energy-momentum conservation as given by (\ref{eq:econs}) and  (\ref{eq:pcons}). The values obtained from this procedure are $\alpha=0.347$, $\delta=0.851$, $\lambda=0.270$ and $\gamma=0.21$.

\begin{figure}[htbp]
\begin{center}
 \includegraphics[angle=0,width=10cm]{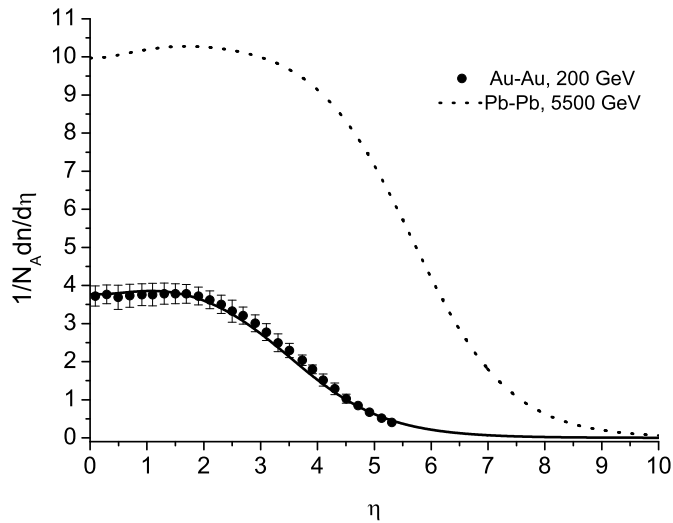}
\caption{ $1/N_A\, dn/d\eta$ in central collisions for Au-Au at $\sqrt s=200$ GeV (full line), data from \cite{Back:2004je}, and prediction for LHC Pb-Pb the at $\sqrt s=5.5$ TeV (dashed line).}
\label{fig:1}
\end{center}
\end{figure}

In Fig. \ref{fig:1} we show the results obtained from (\ref{eq:pseudoforsym}) and the experimental data for RHIC Au-Au central collisions ($\sqrt{s} = 200$ GeV) together with our prediction for the LHC (Pb-Pb central collisions at $\sqrt{s} = 5.5$ TeV. Here, $k$ in the Jacobian (\ref{eq:jac}) was set to fixed value $k=1.37$.

\begin{figure}[htbp]
\begin{center}
 \includegraphics[angle=0,width=10cm]{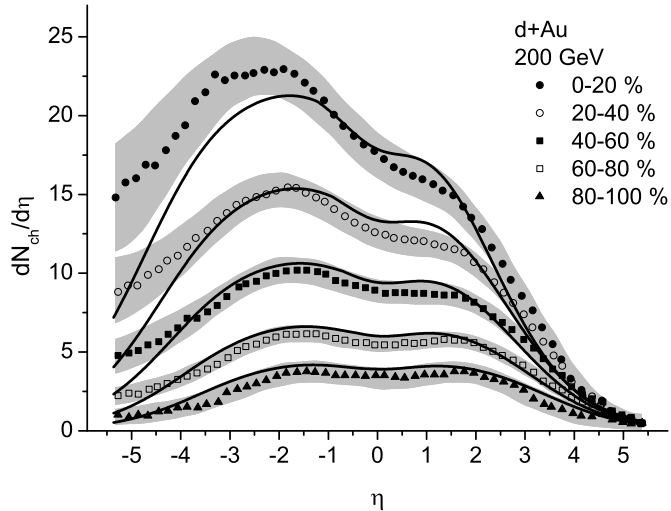}
\caption{$dn/d\eta$ at different centralities from  \cite{Back:2004je} and model curves computed from  (\ref{eq:pseudoforwasym}) and (\ref{eq:pseudobackasym}). Our curves satisfy energy-momentum conservation in the sense of  (\ref{eq:econs}) and  (\ref{eq:pcons}).}
\label{fig:2}
\end{center}
\end{figure}

Fig. \ref{fig:2} we plot the multiplicity distributions for several centrality classes in d-Au collisions at $\sqrt{s}=200$ GeV together with RHIC data \cite{Back:2004je}. The model curves were obtained from (\ref{eq:pseudoforwasym}) and (\ref{eq:pseudobackasym}). Here, $k$ (\ref{eq:jac}) was not set to a fixed value, but rather as a function of the average transverse momentum $p_T^2\rightarrow \langle p_T^2\rangle$ as given by 
(\ref{eq:pt2corrfin}) with $m \rightarrow \langle m\rangle = 0.23$ GeV. The curves fulfil the energy-momentum conservation relations (\ref{eq:econs}) and  (\ref{eq:pcons}).

\begin{figure}[htbp]
\begin{center}
 \includegraphics[angle=0,width=10cm]{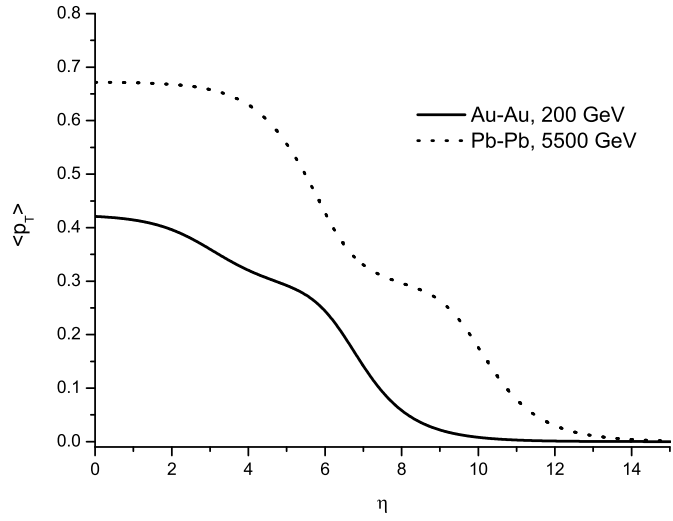}
\caption{Transverse momentum distributions in pseudo-rapidity (\ref{eq:pt2corrfin}) for central Au-Au collisions at $\sqrt{s}=200$ GeV and Pb-Pb at $\sqrt{s}=5.5$ TeV}
\label{fig:3}
\end{center}
\end{figure}

The average transverse momentum $\langle p_T\rangle$ is given, in our model, by (\ref{eq:pt2corrfin}). The single string average transverse momentum is determined from the curves in Fig. \ref{fig:2}, whereas the ratio between the string and proton radius $r/R_P$ was set to the reasonable value of $1/4$. Fig.  \ref{fig:3} shows the average transverse momentum distribution in pseudo-rapidity for both RHIC and LHC symmetric collisions.

A more detailed comparison between the curves obtained from our model and available experimental data both shown in Fig. \ref{fig:2} is given in Table \ref{tab:1}. The number of participant nucleons $N_d$ and $N_{Au}$ for each centrality class
extracted from data \cite{Back:2004je} is compared with the values used in (\ref{eq:pseudoforwasym}) and (\ref{eq:pseudobackasym}) to produce our model curves.
The number of collisions $N_{col}$ extracted from data is compared with that obtained from (\ref{eq:ncol}). Further, The experimentally measured total number of particles with $|\eta|<5.4$ is compared with our model prediction computed from the integral of the curve over the same pseudo-rapidity interval.

\begin{table}[htdp]
\begin{center}
\scalebox{0.8}{
\begin{tabular}{r|rc|rc|rc|rc}
 & 
\multicolumn{2}{c}{$N_d$}  &
\multicolumn{2}{|c}{$N_{Au}$} &
\multicolumn{2}{|c}{$N_{\rm col}$} &
\multicolumn{2}{|c}{$N^{ch}_{|\eta|<5.4}$}  \\
centrality   
& \multicolumn{1}{c}{(exp)}   & (mod) 
& \multicolumn{1}{c}{(exp)}   & (mod) 
& \multicolumn{1}{c}{(exp)}   & (mod) 
& \multicolumn{1}{c}{(exp)}   & (mod) \\ \hline
0-20\%      &$2.0\pm 0.1$ & 1.9    &$13.5\pm 1.0$ & 14.5   & $14.7 \pm 0.9$ & 18.0   &$157\pm 10$  & 143     \\ 
20-40\%     &$1.9\pm 0.1$ & 1.8    &$8.9\pm 0.7$  & 8.2    & $9.8 \pm 0.7$  & 10.0   &$109\pm 7$   & 106     \\ 
40-60\%     &$1.7\pm 0.2$ & 1.5    &$5.4\pm 0.6$  & 4.8    & $5.9\pm 0.6$   & 5.5    &$74\pm 5$    & 74      \\ 
60-80\%     &$1.4\pm 0.2$ & 1.2    &$2.9\pm 0.5$  & 2.4    & $3.1 \pm 0.6$  & 2.6    &$46\pm 3$    & 47      \\ 
80-100\%    &$1.1\pm 0.2$ & 1      &$1.6\pm 0.4$  & 1.2    & $1.7 \pm 0.5$  & 1.2    &$28\pm 3$    & 30   
\end{tabular}
}
\end{center}
\caption{Comparison of d-Au ($\sqrt{s}=200$ GeV) data  \cite{Back:2004je} for different centralities with model. $N_d$, $N_{Au}$ (mod) are those used in (\ref{eq:pseudoforwasym}) and (\ref{eq:pseudobackasym}) to compute the curves shown in Fig. \ref{fig:2}. $N_{col}$ (mod) computed with (\ref{eq:ncol}) using $N_d$, $N_{Au}$ (mod). $N^{ch}_{|\eta|<5.4}$ (mod) computed as the area under the model curves in  Fig. \ref{fig:2}.}
\label{tab:1}
\end{table}%

Table \ref{tab:2} shows the explicit fulfilment of energy-momentum conservation  (\ref{eq:peratio}). The LHS is computed from (\ref{eq:econs}) and (\ref{eq:pcons}) using the model predicted average transverse momentum $\langle p_T^2\rangle$ 
(\ref{eq:pt2corr}) with as before $m\rightarrow \langle p_T\rangle = 0.23$ GeV. The RHS is computed from the values of $N_d$, $N_{Au}$ (mod).

\begin{table}[htdp]
\begin{center}
\scalebox{0.8}{
\begin{tabular}{r|c|c}
 & \multicolumn{2}{|c}{eq.~(\ref{eq:peratio})} \\
centralities  & LHS  & RHS \\ \hline
0-20\%         & 0.76 &0.77 \\
20-40\%        & 0.63 &0.64 \\ 
40-60\%           & 0.52 &0.52 \\ 
60-80\%           & 0.33 &0.33 \\
80-100\%           & 0.09 &0.09 
\end{tabular}
}
\end{center}
\caption{Test of energy-momentum conservation (\ref{eq:peratio})}
\label{tab:2}
\end{table}%

\begin{figure}[htbp]
\begin{center}
 \includegraphics[angle=0,width=6.7cm]{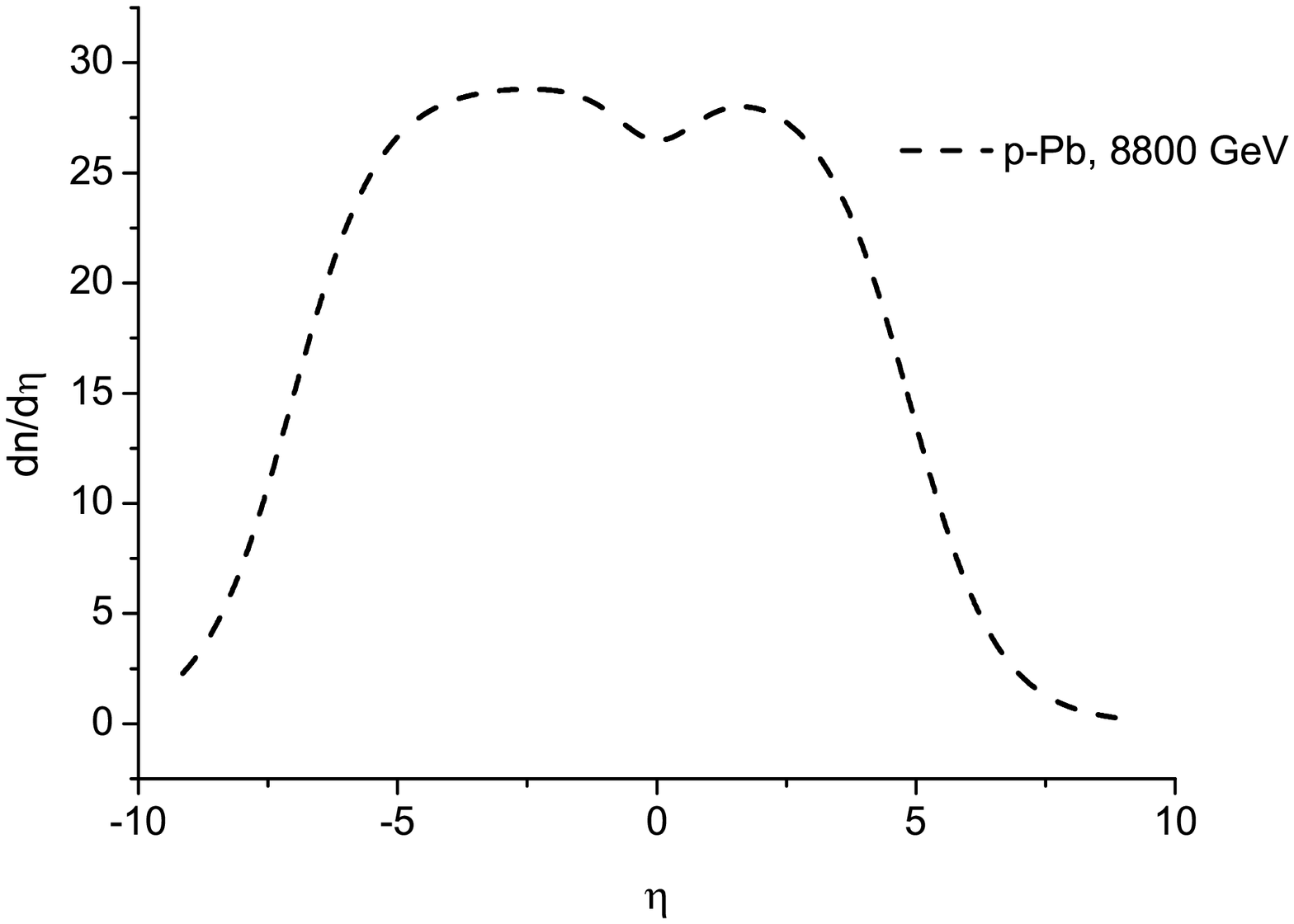}
 \includegraphics[angle=0,width=6.7cm]{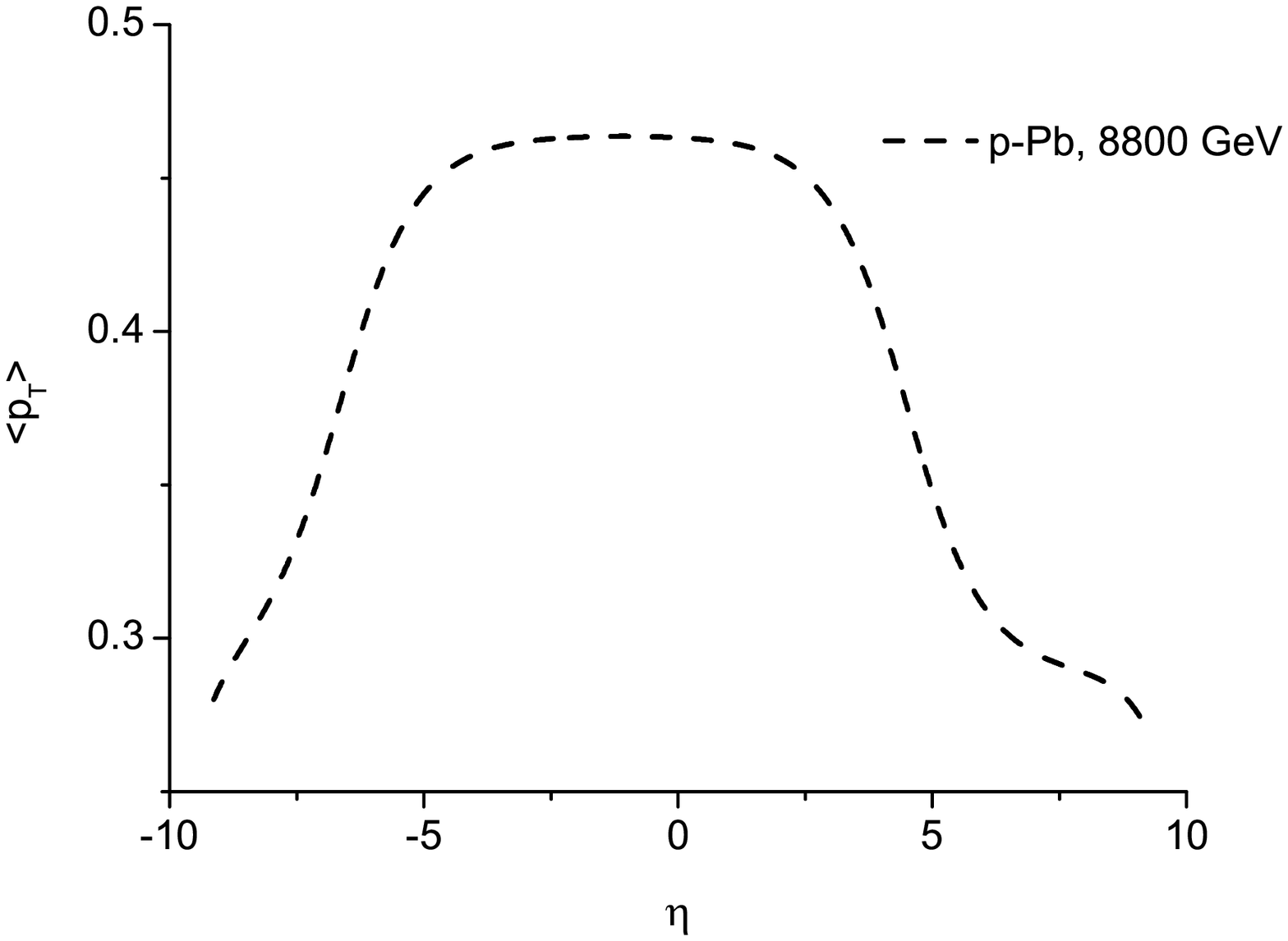}
\caption{Predictions for multiplicity (left) and average transverse momentum (right) distributions as obtained from, respectively,  (\ref{eq:pseudoforwasym}) and (\ref{eq:pseudobackasym}), and (\ref{eq:pt2corrfin})}
\label{fig:4}
\end{center}
\end{figure}

Finally, predictions for both the multiplicity distribution and average transverse momentum for LHC p-Pb collisions ($\sqrt{s}=8.8$ TeV) are shown in Fig. \ref{fig:4}.

\section{Conclusions}
\label{sec:conclusions}

In this paper we were able to present a global description of particle densities and average transverse momentum $\langle p_T\rangle$ for both symmetric and asymmetric collisions for all rapidities, centralities and (high) energies. Although our work was carried out explicitly within percolation theory, we believe that equivalent results should be obtainable in the CGC framework. There, however, the explicit dependence on the number of participants $N_A$ and $N_B$, on the number of collisions $N_{col}$, and control of energy-momentum conservation have not been worked out in detail.

Some of our assumptions still lack full theoretical justification. Namely:

\begin{itemize}
\item 
The relation for $1/N_A \,dn/dy$, (\ref{eq:forsym}) was obtained by taking the limit $\xi \to \infty$ in the dense region, at mid rapidity, and the limit $\xi \to 0$ in the dilute region, at large rapidity. Corrections should be expected at finite $Y$ and $\eta$. The same apllies to the asymmetric formulae (\ref{eq:forwasym}) and (\ref{eq:backasym}).

\item The quantities $(N_A /N_B)^\gamma$ and $(N_B / N_A)^\gamma$, in (\ref{eq:forwasym}) and (\ref{eq:backasym}), representing rescattering corrections necessary if $N_A \neq N_B$, are purely phenomenological without a theoretical ground (Glauber calculus for instance). We do not even know why $\gamma$ (forward) $= \gamma$ (backward) as suggested by the comparison with data.

\item Our formula for $\langle p_T^2\rangle$, (\ref{eq:pt2corr}), describing correctly the kinematical $\langle p_T^2\rangle \eta$ correlation in the $\eta \to \pm \infty$ limits, and making finite and meaningfull the integrals (\ref{eq:econs}) and (\ref{eq:pcons}), is presumably not unique. We noticed that almost 100\% of the integrals (\ref{eq:econs}) and (\ref{eq:pcons}) are contained in an interval $[-\eta,+\eta]$ equal to $[-Y,+Y]$ where $Y$ is the beam rapidity and thus the use of one or another asymptotic formula is not so critical.

These issues clearly warrant future work.

\end{itemize}

\vskip 0.8cm
\noindent{\bf Acknowledgments}
JGM thanks the hospitality of the School of Mathematics, Statistics and Computer Science at the Victoria University of Wellington, its Gravity Group and, in particular, Matt Visser for their kind hospitality. The work of JDD and JGM is partly funded by Funda\c c\~ao para a Ci\^encia e a Tecnologia of Portugal under project CERN/FP/83593/2008.

\end{document}